\documentclass[twocolumn,prb,aps,showpacs,amsmath,amssymb]{revtex4}

\usepackage{graphicx}
\usepackage{dcolumn}
\usepackage{bm}

\begin{document}

\title{Spin channel Keldysh field theory for weakly interacting quantum dots}

\author{Sergey Smirnov}
\email{sergey.smirnov@physik.uni-regensburg.de}
\author{Milena Grifoni}
\affiliation{Institut f\"ur Theoretische Physik, Universit\"at Regensburg,
  D-93040 Regensburg, Germany}

\date{\today}

\begin{abstract}
We develop a low-energy nonequilibrium field theory for weakly interacting
quantum dots. The theory is based on the Keldysh field integral in the spin
channel of the quantum dot described by the single impurity Anderson
Hamiltonian. The effective Keldysh action is a functional of the
Hubbard-Stratonovich magnetization field decoupling the quantum dot spin
channel. We expand this action up to the second order with respect to the
magnetization field, which allows to describe nonequilibrium interacting
quantum dots at low temperatures and weak electron-electron interactions,
up to the contacts-dot coupling energy. Besides its simplicity, an
additional advantage of the theory is that it correctly describes the
unitary limit giving the correct result for the conductance maximum. Thus
our theory establishes an alternative simple method relevant for
investigation of weakly interacting nonequilibrium nanodevices.
\end{abstract}


\maketitle

\section{Introduction}\label{intro}
Nonequilibrium nanoscopic systems having discrete electronic states
\cite{Reed_1988} currently attract unflagging attention of researches from
both experimental and theoretical sides because of practical applications in
various electronic devices. Besides, such systems also provide a unique
platform for fundamental science since they represent a plexus of different
fields of physics leading to new complex and highly nontrivial physical
scenario.

A particularly interesting physics arises when both electron-electron
interactions and nonequilibrium significantly contribute to the state of a
nanoscopic system. The system's differential conductance, as is well known,
may then enhance \cite{Glazman_1988,Ng_1988} and exceed the value it would
have without electronic correlations. This enhancement, taking place at low
temperatures, signifies appearance of new physics due to the system's
transition into a resonant many-particle Kondo regime discovered first in
the context of magnetic alloys \cite{de_Haas_1934,Kondo_1964,Hewson_1997}.

The single-impurity Anderson model (SIAM) \cite{Anderson_1961} is one of the
main theoretical paradigms which is able to capture basic physics of
nonequilibrium interacting nanoscopic systems. It describes a quantum dot
(QD) with a single spin-degenerate level coupled to two fermionic contacts.
The contacts have different chemical potentials with the difference
specifying the voltage applied to the QD. This voltage is the source of
nonequilibrium.

Quantum transport theories built upon SIAM can be divided into two classes:
1) operator based theories and 2) field integral based theories. Within the
first class one directly uses the second quantized operators while within the
second class one transforms these operators into fields whose dynamics is
governed by a certain effective action.

Among numerous examples of the first class theories are perturbation theories
in the electron-electron interaction \cite{Fujii_2003,Muehlbacher_2011} as
well as in the tunneling amplitude \cite{Sivan_1996}, noncrossing
approximation \cite{Meir_1993,Wingreen_1994,Hettler_1995}, equations of
motion \cite{Meir_1993,Entin_2005,Roermund_2010}, mean-field approximation
\cite{Aguado_2000,Lopez_2003}, renormalization group theories \cite{Gezzi_2007,
Bulla_2008,Anders_2008,Heidrich_2009}. At the same time the relatively new
second class is not so wide since field integral concepts in physics of
nonequilibrium interacting nanoscopic systems are just on the way of growing
emergence. Here examples are given by analytical \cite{Ratiani_2009,
Smirnov_2011,Smirnov_2011a} and numerical \cite{Weiss_2008,Eckel_2010} Keldysh
field integral theories.

The Anderson impurity model has two distinct fixed points, the weak coupling
fixed point and the strong coupling or Kondo fixed point, each one being a
Fermi liquid \cite{Altland_2010}.

Analytical field integral oriented theories are mainly based on slave-particle
\cite{Barnes_1976,Coleman_1984,Kotliar_1986,Coleman_1987,Zou_1988} strong
coupling fixed point approaches. For example in Ref. \onlinecite{Ratiani_2009}
the saddle point analysis is applicable at temperatures below the Kondo
temperature $T_\text{K}$ and thus the unitary limit is within its temperature
range. However, being a $1/N$ expansion it gives an incorrect value of the
conductance maximum for spin-$1/2$. In Refs.
\onlinecite{Smirnov_2011,Smirnov_2011a} the effective Keldysh action is expanded
around the zero slave-bosonic field configuration up to the second order in the
slave-bosonic fields and this restricts those theories to temperature range
close to and above $T_\text{K}$. These examples show that either the unitary limit
is incorrectly described quantitatively or it cannot be reached at all due to
temperature limitations of theories. However, it is desirable to have a field
integral theory treating the unitary limit properly since this limit gets more
and more feasible in modern experiments \cite{Kretinin_2011} both for the strong
coupling and weak coupling fixed points regimes.

A practical guide for developing a second class theory having a proper treatment
of the unitary limit in the weak coupling fixed point regime is given by the first
class theories, namely perturbation theories being expansions in powers of the
electron-electron interaction. Indeed, these theories
\cite{Fujii_2003,Muehlbacher_2011} are applicable at zero temperature and reproduce
the correct value of the conductance maximum, $2e^2/h$. This gives one the cue that
in the context of the field integration a theory valid at zero temperature and
having the correct unitary limit might be obtained through the expansion of its
effective action in powers of the electron-electron interaction. Of course, such an
expansion of the effective action means also an expansion in powers of a certain
field. This field turns out to be non-unique and its choice is not obvious a priori.
At this stage one usually relies upon various physical motivations which could
simplify mathematical formulation and achieve physical clarity.

In this paper we choose this field as the Hubbard-Stratonovich field
decoupling the electronic correlations in their spin channel. This means that
such a magnetization field is sensitive to the QD spin fluctuations induced by
the electron-electron interaction. Since it has a magnetic origin it is also
susceptible to the QD magnetic properties. In particular, when the magnetic
symmetry is violated, {\it e.g.}, in the presence of a magnetic field, either
directly applied to the QD or indirectly induced in the QD by the ferromagnetic
contact proximity effect, the minimum of the effective Keldysh action moves from
the zero magnetization field configuration and the new extremum provides the
effective magnetic field experienced by the QD electron dynamics. On the contrary,
in the absence of any magnetic structure the effective Keldysh action simplifies
admitting only even powers of the magnetization field.

In general the effective Keldysh action is a nonlinear functional of the
magnetization field. Here we expand it up to the second order in this field,
which is also a second order expansion in the electron-electron interaction. The
quadratic model is an expansion about the weak coupling fixed point, where the
saddle-point magnetization vanishes. Thus such a theory must reproduce the unitary
limit because it is an expansion about a Fermi liquid fixed point. Therefore, the
goal of the present research is to develop a quadratic spin channel Keldysh field
integral formalism to provide an alternative theoretical tool for investigation of
weakly correlated nonequilibrium nanosystems.

The paper is organized as follows. Section \ref{qdsc} introduces the spin
channel in the single impurity Anderson Hamiltonian while Section \ref{sckfi}
converts it into the Keldysh field integral framework and provides the general
form of the effective Keldysh action as a functional of the Hubbard-Stratonovich
classical and quantum magnetization fields. In Section \ref{scekatdos} this
action is expanded up to the second order in the magnetization fields and
afterwards it is used to obtain the QD tunneling density of states. Finally, the
results are shown in Section \ref{res} and with Section \ref{concl} we conclude.

\section{Quantum dot spin channel}\label{qdsc}
We first formulate the problem on the operator level and prepare at this stage
for its subsequent field integral formulation in the QD spin channel.

The single impurity Anderson Hamiltonian reads,
\begin{equation}
\hat{H}_\text{d}=\sum_{\sigma}\epsilon_\text{d}\hat{n}_{\text{d},\sigma}+U\hat{n}_{\text{d},\uparrow}\hat{n}_{\text{d},\downarrow},
\label{siam_ham}
\end{equation}
where $\sigma=\uparrow,\downarrow$, $\hat{n}_{\text{d},\sigma}=d^\dagger_\sigma d_\sigma$,
$\{d^\dagger_\sigma,\,d_\sigma\}$ are the QD creation and annihilation electronic
operators, $\epsilon_\text{d}$ is the QD energy level and $U$ is the strength of
the electron-electron interaction in the QD.

The contacts are fermionic noninteracting reservoirs described by the following
Hamiltonian:
\begin{equation}
\hat{H}_\text{C}=\sum_a\epsilon_ac^\dagger_ac_a,
\label{cont_ham}
\end{equation}
where $a=\{x,k,\sigma\}$ is the contact set of quantum numbers including the contacts
label, $x=\text{L,R}$ (left and right contacts), $\{c^\dagger_a,\,c_a\}$ are the contacts
creation and annihilation operators and $\epsilon_a$ identifies the contacts
single-particle energies. The contacts are in equilibrium described by the Fermi-Dirac
distributions, $n(\epsilon)=\{\exp[(\epsilon-\mu_x)/kT]+1\}^{-1}$, where $\mu_x$ are the
contacts chemical potentials, defining the voltage applied to the QD as
$V\equiv(\mu_\text{R}-\mu_\text{L})/e$, and $T$ is the contacts temperature which is
assumed to be the same in the left and right contacts.

The QD and contacts interact through a tunneling coupling given by the tunneling
Hamiltonian,
\begin{equation} 
\hat{H}_\text{T}=\sum_{a\sigma}(c^\dagger_aT_{a\sigma}d_\sigma+d^\dagger_\sigma T^*_{a\sigma}c_a),
\label{tun_ham}
\end{equation}
where $T_{a\sigma}$ are the tunneling matrix elements.

In order to construct a field integral in the QD spin channel one has to rewrite the
QD Hamiltonian in such a way that the coupling to the QD electron spin variable
becomes apparent. This can be achieved, {\it e.g.}, using the following equality
\begin{equation}
\hat{n}_{\text{d},\uparrow}\hat{n}_{\text{d},\downarrow}=\frac{1}{2}(\hat{n}_{\text{d},\uparrow}+\hat{n}_{\text{d},\downarrow})-
\frac{1}{2}(\hat{n}_{\text{d},\uparrow}-\hat{n}_{\text{d},\downarrow})^2.
\label{sc_eq}
\end{equation}
As a result the QD Hamiltonian acquires the form explicitly involving the QD electron
spin degree of freedom,
\begin{equation}
\hat{H}_\text{d}=\sum_\sigma\biggl(\epsilon_\sigma+\frac{U}{2}\biggl)\hat{n}_{\text{d},\sigma}-\frac{U}{2}\biggl(\sum_\sigma\sigma\hat{n}_{\text{d},\sigma}\biggl)^2.
\label{siam_sc_ham}
\end{equation}

The QD Hamiltonian in the form of Eq. (\ref{siam_sc_ham}) together with Eqs.
\ref{cont_ham} and \ref{tun_ham} constitute a nonequilibrium interacting problem
with the full Hamiltonian $\hat{H}=\hat{H}_\text{d}+\hat{H}_\text{C}+\hat{H}_\text{T}$.
The explicit presence of the QD electron spin in the operator formulation allows
one to introduce within the Keldysh field integral framework classical and quantum
fields directly connected to the QD spin channel dynamics, as it is shown in the
next section.

\section{Spin channel Keldysh field integral}\label{sckfi}
An equality similar to Eq. (\ref{sc_eq}) has been utilized
\cite{Hertz_1976,Altland_2010} to explore quantum critical phenomena, in particular
itinerant magnetic phases, using a field integral in the imaginary (or Matsubara)
time formulation. The field integral in that approach is obtained by integrating
out the fermionic degrees of freedom and obtaining an effective action as a
functional of the Hubbard-Stratonovich field decoupling the spin channel. It turns
out that such a Hubbard-Stratonovich field has a physical meaning of magnetization
and it is sensitive to magnetic properties of systems.

In the same spirit, using real time and integrating out the fermionic degrees of
freedom, one arrives at the Keldysh field integral \cite{Altland_2010} for SIAM in
the QD spin channel.

Here before integrating out the fermionic degrees of freedom the action is identical
to the one in Eq. (6) of Ref. \onlinecite{Weiss_2008}. However, after that stage we
perform the Keldysh rotation \cite{Altland_2010} and, instead
of Ising-like discrete spin fields, we use a continuous Hubbard-Stratonovich field
from Refs. \onlinecite{Hertz_1976,Altland_2010}.

One of the main QD physical observables is the tunneling density of states (TDOS),
$\nu_\sigma(\epsilon)\equiv-(1/\hbar\pi)\text{Im}[G^+_{\text{d}\,\sigma\sigma}(\epsilon)]$
($G^+_{\text{d}\,\sigma\sigma}(\epsilon)$ is the QD retarded Green's function; below the
upper indices $+$ and $-$ always denote, respectively, the retarded and advanced
components of matrices in the Keldysh space), with the corresponding Keldysh field
integral representation,
\begin{widetext}
\begin{equation}
\begin{split}
\nu_\sigma(\epsilon)=\frac{1}{2\pi\hbar}\int dt\exp\biggl(\frac{i}{\hbar}\epsilon t\biggl)
\int\mathcal{D}[m(t)]\exp\biggl\{\frac{i}{\hbar}S_\text{eff}[m(t)]\biggl\}
\biggl\{G^+(\sigma t|\sigma 0)-G^-(\sigma t|\sigma 0)\biggl\},
\end{split}
\label{tdos_sc_kfi}
\end{equation}
\begin{equation}
S_\text{eff}[m(t)]=-\int dt Um_c(t)m_q(t)-
i\hbar\,\text{tr}\bigl\{\ln[G^{-1}(\alpha t|\alpha't')]-\ln[G^{(0)-1}(\alpha t|\alpha' t')]\bigl\},
\label{eff_a}
\end{equation}
\begin{equation}
G^{-1}(\alpha t|\alpha't')=-
\begin{pmatrix}
iG_\text{d}^{(0)-1}(\sigma t|\sigma't')-\frac{i}{\hbar}\sigma U M_\text{HS}(\sigma t|\sigma't')&\frac{i}{\hbar}M^\dagger_\text{T}(\sigma t|a't')\\
\frac{i}{\hbar}M_\text{T}(a t|\sigma't')&iG_\text{C}^{(0)-1}(at|a't')
\end{pmatrix}.
\label{M_M0_mtr}
\end{equation}
\end{widetext}
Here $m_c(t)$, $m_q(t)$ are the classical and quantum magnetization fields being the
Hubbard-Stratonovich fields decoupling the QD spin channel and
$G^{(0)-1}(\alpha t|\alpha't')=G^{-1}(\alpha t|\alpha't')$ with $U=0$. In Eq.
(\ref{M_M0_mtr}) $G_\text{d}^{(0)-1}(\sigma t|\sigma't')$, $G_\text{C}^{(0)-1}(at|a't')$,
$M_\text{HS}(\sigma t|\sigma't')$ and $M_\text{T}(a t|\sigma't')$ are the following
matrices in the Keldysh space:
\begin{widetext}
\begin{equation}
G_\text{d}^{(0)-1}(\sigma t|\sigma't')\equiv\delta_{\sigma\sigma'}
\begin{pmatrix}
\bigl[i\frac{\partial}{\partial t}-\frac{\epsilon_\text{d}+U/2}{\hbar}+i0^+\bigl]\delta(t-t')&i0^+f_\text{d}(t-t')\\
0&\bigl[i\frac{\partial}{\partial t}-\frac{\epsilon_\text{d}+U/2}{\hbar}-i0^+\bigl]\delta(t-t')
\end{pmatrix},
\label{Gd0m1}
\end{equation}
\begin{equation}
G_\text{C}^{(0)-1}(at|a't')\equiv\delta_{aa'}
\begin{pmatrix}
\bigl[i\frac{\partial}{\partial t}-\frac{\epsilon_a}{\hbar}+i0^+\bigl]\delta(t-t')&i0^+f_a(t-t')\\
0&\bigl[i\frac{\partial}{\partial t}-\frac{\epsilon_a}{\hbar}-i0^+\bigl]\delta(t-t')
\end{pmatrix},
\label{GC0m1}
\end{equation}
\begin{equation}
M_\text{HS}(\sigma t|\sigma't')\equiv\delta_{\sigma\sigma'}\delta(t-t')
\begin{pmatrix}
m_c(t)&\frac{1}{2}m_q(t)\\
\frac{1}{2}m_q(t)&m_c(t)
\end{pmatrix},\quad
M_\text{T}(at|\sigma't')\equiv\delta(t-t')
\begin{pmatrix}
T_{a\sigma}&0\\
0&T_{a\sigma}
\end{pmatrix},
\label{MHS_MT}
\end{equation}
\end{widetext}
where $f_\text{d}(t)$ and $f_a(t)$ are the Fourier transforms of the QD and contacts
distribution functions, respectively.

\section{Spin channel effective Keldysh action and TDOS}\label{scekatdos}
The effective Keldysh action, Eq. (\ref{eff_a}), is a nonlinear functional of the
magnetization fields $m_c(t)$ and $m_q(t)$. In this section we want to investigate
which kind of physics is described by this action when it is expanded up to the
second order in the magnetization fields.

In this paper we are only interested in the effective field theory for QDs in the
absence of any magnetic structure. It is easy to see that in this case Eq.
(\ref{eff_a}) does not have odd powers in the magnetization fields. Indeed, the
absence of any spin dependence just results in traces of the traceless Pauli
operators $\hat{\sigma}_z$ eliminating in this way all odd powers of the
magnetization fields from Eq. (\ref{eff_a}) implying that in this case
$S_\text{eff}[-m(t)]=S_\text{eff}[m(t)]$.

Therefore, the second order expansion of the effective Keldysh action cannot have
linear terms. Since the zero order term is equal to zero, the only nonvanishing
terms in this expansion are terms of the second order in the magnetization fields.
Performing the expansion of the functional $S_\text{eff}[m(t)]$ one finds,
\begin{equation}
\begin{split}
&\frac{i}{\hbar}S_\text{eff}[m(t)]=-\int\frac{d\omega}{2\pi}
\begin{bmatrix}m_c(-\omega)&m_q(-\omega)\end{bmatrix}\times\\
&\times\begin{pmatrix}
0&\frac{iU}{2\hbar}+\frac{U^2}{\hbar^2}\Sigma^-_V(\omega)\\
\frac{iU}{2\hbar}+\frac{U^2}{\hbar^2}\Sigma^+_V(\omega)&\frac{U^2}{\hbar^2}\Sigma^\text{K}_V(\omega)
\end{pmatrix}
\begin{bmatrix}
m_c(\omega)\\m_q(\omega)
\end{bmatrix},
\end{split}
\label{eff_a_so}
\end{equation}
where $m_c(\omega)$ and $m_q(\omega)$ are the Fourier transforms of the classical
and quantum magnetization fields and $\Sigma^{+,-,\text{K}}_V(\omega)$ are the retarded,
advanced and Keldysh components of the self-energy matrix. Assuming a symmetric
energy independent spin diagonal QD-contacts coupling,
$T_{a\sigma'}=\delta_{\sigma\sigma'}\mathcal{T}$, and an energy independent contacts
density of states, $\nu_\text{C}$, we find the following analytical expressions for
$\Sigma^{+,-,\text{K}}_V(\omega)$ ($\Sigma^-_V(\omega)=[\Sigma^+_V(\omega)]^*$):
\begin{equation}
\begin{split}
&\Sigma^+_{V}(\omega)=\sum_{s,s'=\{+,-\}}I^+(s\omega,s'V),\\
&\Sigma^\text{K}_{V}(\omega)=\sum_{s=\{+,-\}}[I_1^\text{K}(\omega,sV)+I_2^\text{K}(\omega,sV)],
\end{split}
\label{self_en}
\end{equation}
\begin{equation}
\begin{split}
&I^+(\omega,V)=\frac{1}{4\pi}\frac{\Gamma}{i\omega(2\Gamma-i\hbar\omega)}\times\\
&\times\biggl[i\frac{\hbar\omega}{\Gamma}\psi(x^+_2)+2\biggl(1-i\frac{\hbar\omega}{2\Gamma}\biggl)\psi(x^+_1)-2\psi(y^+_1)\biggl]-\\
&-\frac{i}{4}\frac{e^z-1}{e^z+1}\frac{\hbar}{2\Gamma-i\hbar\omega},
\end{split}
\label{I+}
\end{equation}
\begin{equation}
\begin{split}
&I^\text{K}_1(\omega,V)=-i\text{Im}\biggl\{\frac{1}{8\pi}\frac{\Gamma}{i\omega(2\Gamma+i\hbar\omega)}\times\\
&\times\coth\biggl(\frac{\hbar\omega}{2kT}\biggl)\biggl[i\frac{\hbar\omega}{\Gamma}\bigl(\psi(x^+_2)-\psi(y^+_2)\bigl)-\\
&-2\biggl(1+i\frac{\hbar\omega}{2\Gamma}\biggl)\bigl(\psi(x^+_1)-\psi(y^+_1)\bigl)+2\bigl(\psi(y)-\\
&-\psi(x^+_1)\bigl)\biggl]-\frac{i}{4}\frac{e^z+e^p}{(e^z+1)(e^p+1)}\frac{\hbar}{2\Gamma+i\hbar\omega}\biggl\},
\end{split}
\label{IK1}
\end{equation}
\begin{equation}
\begin{split}
&I^\text{K}_2(\omega,V)=-i\text{Im}\biggl\{\frac{1}{8\pi}\frac{\Gamma}{i\omega(2\Gamma+i\hbar\omega)}\times\\
&\times\coth\biggl(\frac{\hbar\omega+eV}{2kT}\biggl)\biggl[i\frac{\hbar\omega}{\Gamma}\bigl(\psi(x^+_2)-\psi(x^-_2)\bigl)-\\
&-2\biggl(1+i\frac{\hbar\omega}{2\Gamma}\biggl)\bigl(\psi(x^+_1)-\psi(y^-_1)\bigl)+2\bigl(\psi(y)-\\
&-\psi(x^-_1)\bigl)\biggl]-\frac{i}{4}\frac{e^z+e^q}{(e^z+1)(e^q+1)}\frac{\hbar}{2\Gamma+i\hbar\omega}\biggl\},
\end{split}
\label{IK2}
\end{equation}
where $\Gamma\equiv\pi\nu_\text{C}|\mathcal{T}|^2$, $\psi(x)$ is the digamma
function and $x^+_{1,2}\equiv1/2+(i/2\pi kT)(\epsilon_\text{d}+U/2+eV/2)\pm\Gamma/2\pi kT$, 
$x^-_{1,2}\equiv1/2+(i/2\pi kT)(\epsilon_\text{d}+U/2-eV/2)\pm\Gamma/2\pi kT$,
$y^{\pm}_{1,2}\equiv x^{\pm}_{1,2}-i\hbar\omega/2\pi kT$,
$y\equiv x^+_1+i\hbar\omega/2\pi kT$,
$z\equiv(\epsilon_\text{d}+U/2+eV/2)/kT+i\Gamma/kT$,
$p\equiv z-\hbar\omega/kT,\quad q\equiv z-\hbar\omega/kT-eV/kT$.

With the effective Keldysh action (\ref{eff_a_so}) one obtains the following
expression for the QD TDOS:
\begin{equation}
\begin{split}
&\nu_\sigma(\epsilon)=\nu_0(\epsilon)+\frac{iU^2}{4\pi^2\hbar^3}\biggl\{\biggl[D^+\biggl(\frac{\epsilon}{\hbar}\biggl)\biggl]^2\times\\
&\times\int\frac{d\omega}{2\pi}\biggl[J^\text{K}_V\biggl(\frac{\epsilon}{\hbar}-\omega\biggl)D^+(\omega)+\\
&+\frac{1}{2}J^+_V\biggl(\frac{\epsilon}{\hbar}-\omega\biggl)D^\text{K}_V(\omega)\biggl]-\biggl[D^-\biggl(\frac{\epsilon}{\hbar}\biggl)\biggl]^2\times\\
&\times\int\frac{d\omega}{2\pi}\biggl[J^\text{K}_V\biggl(\frac{\epsilon}{\hbar}-\omega\biggl)D^-(\omega)+\\
&+\frac{1}{2}J^-_V\biggl(\frac{\epsilon}{\hbar}-\omega\biggl)D^\text{K}_V(\omega)\biggl]+\frac{1}{2}D^\text{K}_V\biggl(\frac{\epsilon}{\hbar}\biggl)\times\\
&\times\biggl[D^+\biggl(\frac{\epsilon}{\hbar}\biggl)-D^-\biggl(\frac{\epsilon}{\hbar}\biggl)\biggl]\int\frac{d\omega}{2\pi}\biggl[J^+_V\biggl(\frac{\epsilon}{\hbar}-\omega\biggl)\times\\
&\times D^-(\omega)+J^-_V\biggl(\frac{\epsilon}{\hbar}-\omega\biggl)D^+(\omega)\biggl]\biggl\},
\end{split}
\label{tdos_sc_fin}
\end{equation}
where
\begin{equation}
\nu_0(\epsilon)=\frac{1}{\pi}\frac{\Gamma}{[\epsilon-(\epsilon_\text{d}+U/2)]^2+\Gamma^2},
\label{tdos_0}
\end{equation}
\begin{equation}
\begin{split}
&J^\text{K}_V(\omega)\equiv\frac{-i\hbar^2\pi\Sigma^\text{K}_V(\omega)}{[\hbar/2-U\Sigma^+_V(\omega)][\hbar/2-U\Sigma^-_V(\omega)]},\\
&J^\pm_V(\omega)\equiv\frac{-i\hbar^2\pi/U}{\hbar/2-U\Sigma^\pm_V(\omega)},
\end{split}
\label{J+-K}
\end{equation}
\begin{equation}
\begin{split}
&D^{\pm}(\omega)\equiv\frac{\hbar}{\hbar\omega-(\epsilon_\text{d}+U/2)\pm i\Gamma},\\
&D^\text{K}_V(\omega)\equiv\frac{-i\hbar\Gamma\sum_xF_x(\omega,V)}{[\hbar\omega-(\epsilon_\text{d}+U/2)]^2+\Gamma^2},
\end{split}
\label{D+-K}
\end{equation}
and $F_\text{L,R}(\omega,V)\equiv\tanh[(\hbar\omega\pm eV/2)/2kT]$.

\section{Results}\label{res}
Using Eq. (\ref{tdos_sc_fin}) one can obtain the QD TDOS using a numerical
frequency integration. Since the expansion (\ref{eff_a_so}) of the
effective Keldysh action (\ref{eff_a}) in the classical, $m_c(t)$, and
quantum, $m_q(t)$, magnetization fields is an expansion about the weak
coupling fixed point, our simple quadratic field integral theory is valid
only for weakly interacting QDs, $U\lesssim\Gamma$. Such a theory must
reproduce at low temperatures the correct value $2e^2/h$ of the conductance
maximum, known as the unitary limit. Let us recall that this is not the case
in existing Keldysh field integral strong coupling fixed point theories both
analytical \cite{Ratiani_2009,Smirnov_2011,Smirnov_2011a} and numerical
\cite{Weiss_2008,Eckel_2010}. In the analytical theories the unitary limit
is either incorrectly described quantitatively \cite{Ratiani_2009} or it
cannot be reached at all due to temperature limitations related to
proliferation of slave-bosonic oscillations
\cite{Smirnov_2011,Smirnov_2011a}. In the numerical theories
\cite{Weiss_2008,Eckel_2010} the unitary limit is difficult to reach because
the memory time becomes infinite at zero temperature.
\begin{figure}
\includegraphics[width=7.6 cm]{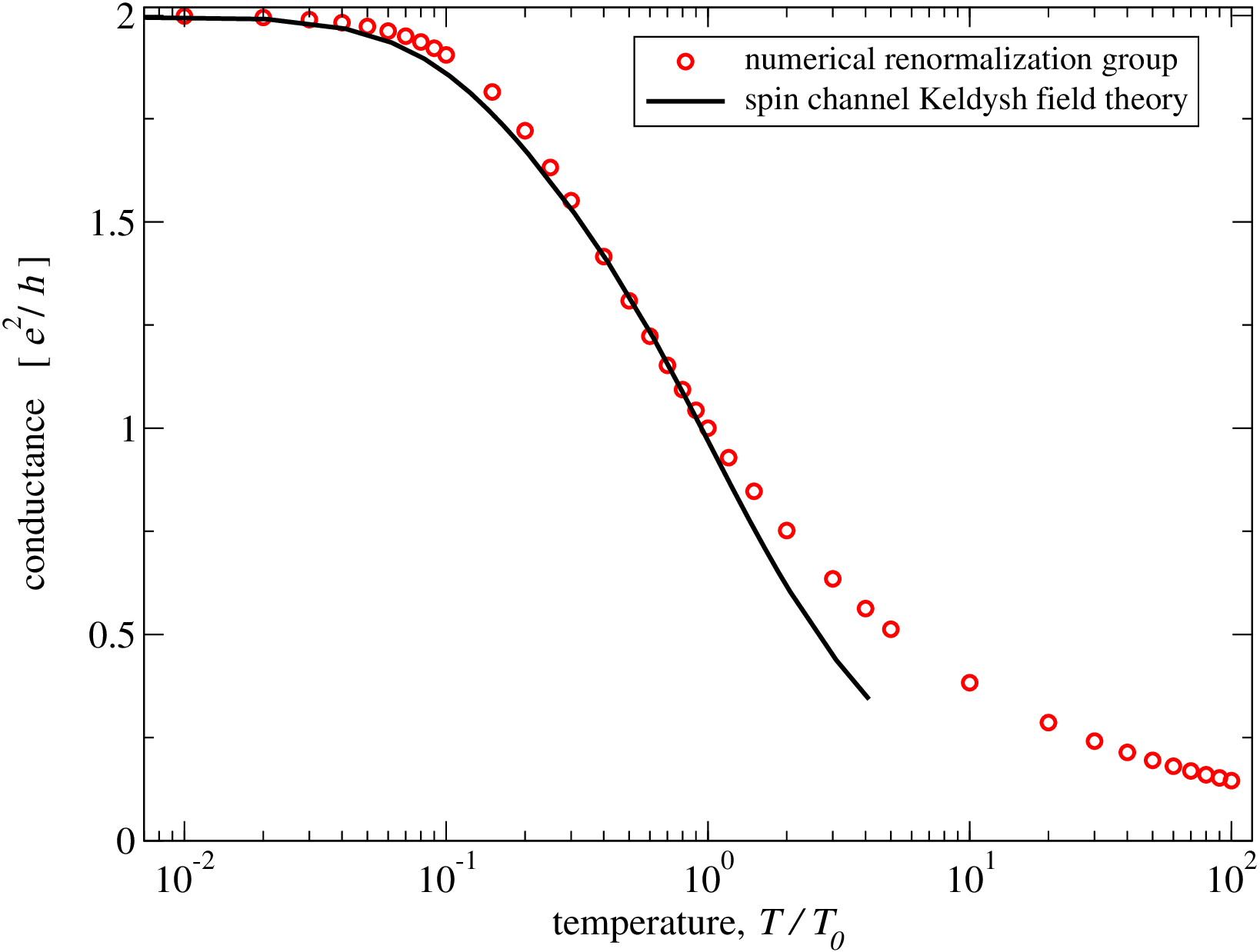}
\caption{\label{figure_1} (Color online) Temperature dependence of the
differential conductance maximum at the symmetric point obtained from the
present spin channel Keldysh field integral theory for $U=0.9\,\Gamma$,
$\epsilon_\text{d}=U/2$. Here $kT_0$ is the zero temperature QD TDOS half
width at half maximum. The red circles show the universal temperature
dependence of the differential conductance maximum obtained in the
numerical renormalization group theory
\cite{Goldhaber-Gordon_1998a,Grobis_2008}. In this case $kT_0$ is the Kondo
temperature which is approximately equal to the zero temperature QD TDOS half
width at half maximum.}
\end{figure}
\begin{figure}
\includegraphics[width=7.6 cm]{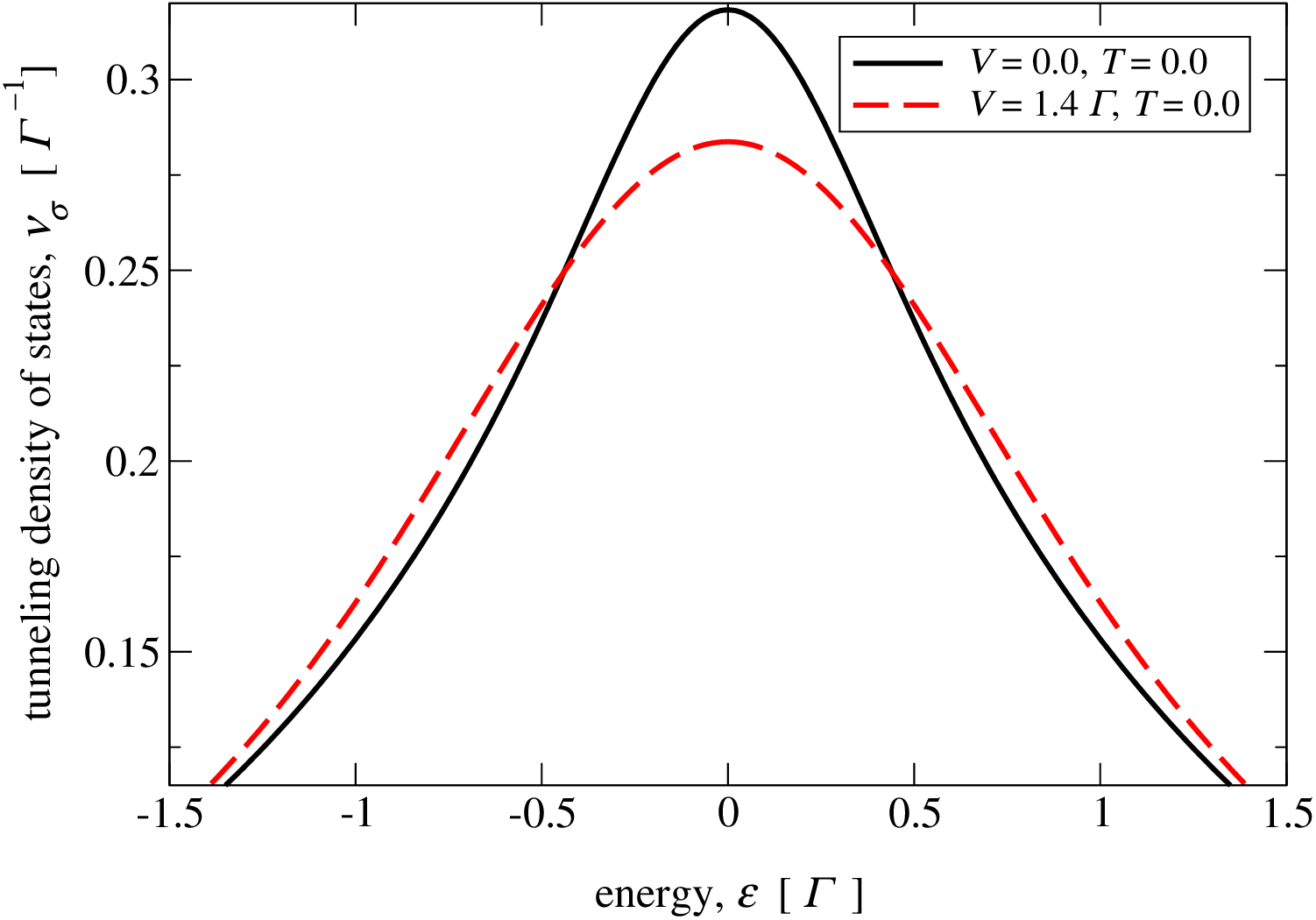}
\caption{\label{figure_2} (Color online) Equilibrium and nonequilibrium QD
  TDOS (\ref{tdos_sc_fin}) at zero temperature, $U=0.8\,\Gamma$,
  $\epsilon_\text{d}=U/2$. The effect of a finite voltage is to decrease and
  broaden the QD TDOS.}
\end{figure}

In Fig. \ref{figure_1} we show the temperature dependence of the differential
conductance maximum. This figure confirms the consistency  of the results
obtained in the previous section. Indeed, at low temperatures they give the
correct value of the differential conductance maximum, $2e^2/h$, as it must
be for the expansion about the weak coupling Fermi liquid fixed point.
Additionally we plot the universal temperature dependence of the differential
conductance maximum obtained in the numerical renormalization group theory.
The comparison between the curves demonstrates that when $U$ increases so that
at the symmetric point the system becomes closer to the Kondo regime the low
temperature behavior of the differential conductance maximum obtained from the
spin channel Keldysh field integral theory becomes closer to the universal
temperature dependence of the differential conductance maximum obtained in the
numerical renormalization group theory.
\begin{figure}
\includegraphics[width=7.6 cm]{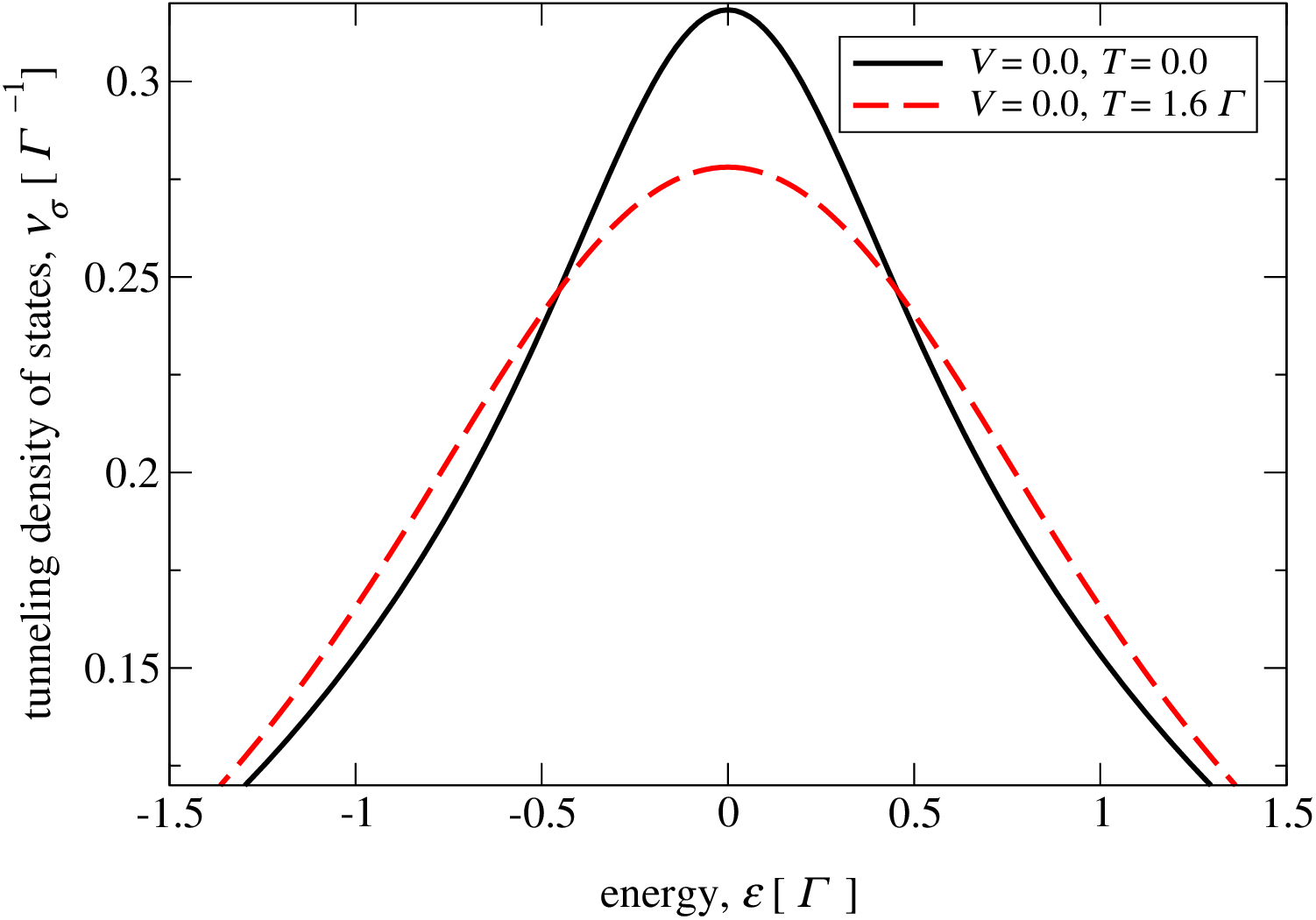}
\caption{\label{figure_3} (Color online) Equilibrium QD TDOS
  (\ref{tdos_sc_fin}) at zero and finite temperatures, $U=0.8\,\Gamma$,
  $\epsilon_\text{d}=U/2$. As one can see the effect of a finite voltage in
  Fig. \ref{figure_2} is similar to the effect of a finite temperature in this
  figure.}
\end{figure}
\begin{figure}
\includegraphics[width=7.6 cm]{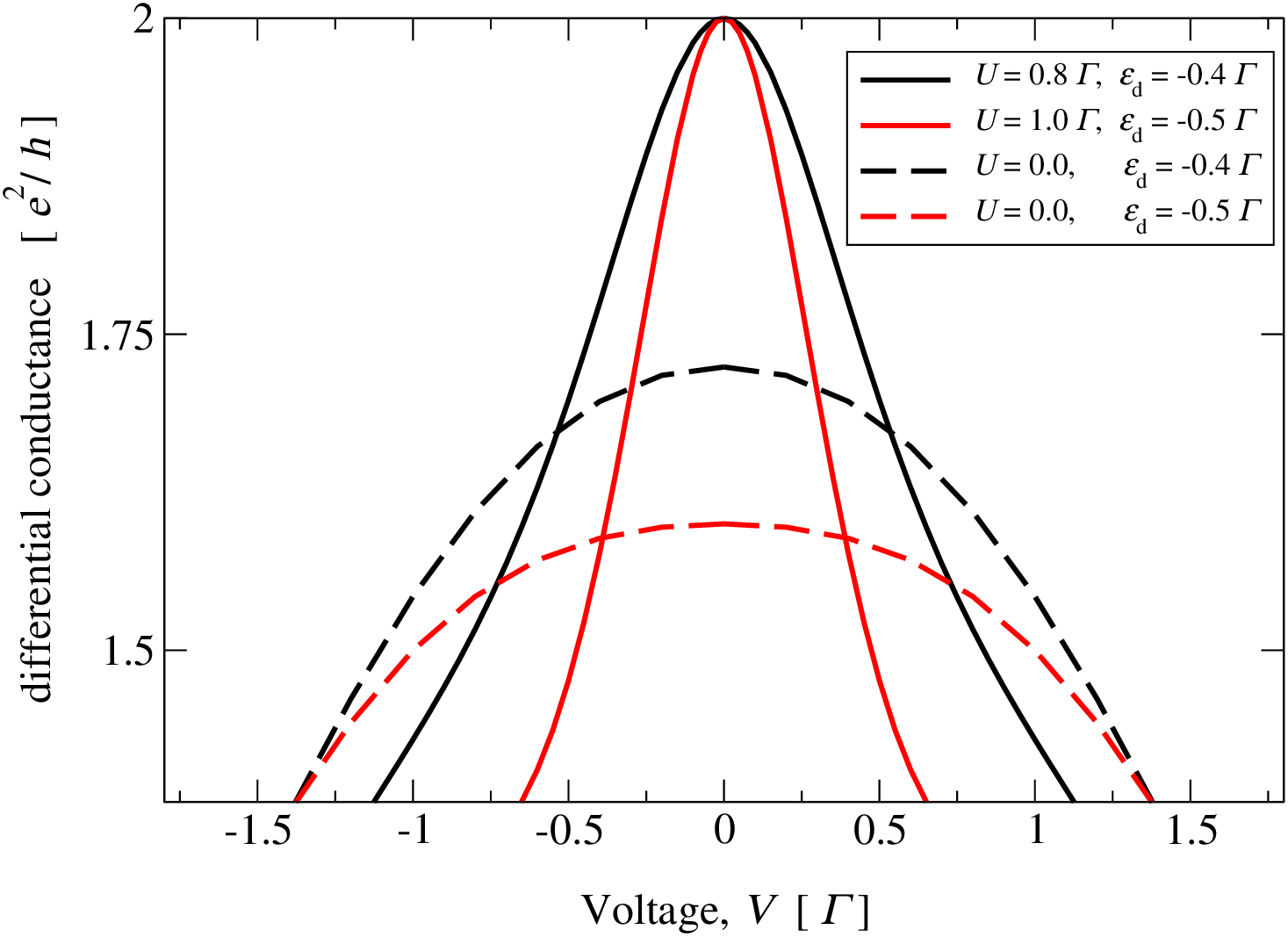}
\caption{\label{figure_4} (Color online) Zero temperature QD differential
  conductance as a function of the applied voltage. Both interacting and
  noninteracting cases are shown. Due to the electron-electron interaction
  the QD is in the resonant many-particle state where its differential
  conductance is enhanced at low voltages in comparison with the
  noninteracting counterpart. The maximum is equal to the correct value
  $2e^2/h$.}
\end{figure}

The results of our spin channel Keldysh field integral theory show that in the
weak coupling fixed point regime both finite voltages and finite temperatures
have a similar impact on the QD TDOS making it lower and broader in comparison
with the zero temperature equilibrium QD TDOS as one can see from Figs.
\ref{figure_2} and \ref{figure_3}. This behavior is different from the one in
the strong coupling fixed point regime where the finite voltage splits the
Kondo resonance as soon as it becomes bigger than its width
\cite{Fujii_2003,Wingreen_1994,Smirnov_2011,Smirnov_2011a}.

Finally, the quadratic spin channel Keldysh field integral theory can also be
used to calculate the QD differential conductance as a function of the applied
voltage. In Fig. \ref{figure_4} the zero temperature differential conductance
is shown for the noninteracting, $U=0$, and interacting, $U=0.8\Gamma$,
$U=1.0\Gamma$, QDs. Once again, as in Fig. \ref{figure_1}, the correct value of
the maximum in Fig. \ref{figure_4} proves the consistency of the quadratic spin
channel Keldysh field integral theory.

\section{Conclusion}\label{concl}
We have developed a spin channel Keldysh field integral theory for
nonequilibrium interacting QDs. To describe nonequilibrium interacting states
of the QD we have introduced a collective degree of freedom, a magnetization
field, being the Hubbard-Stratonovich field decoupling the spin channel of the
electron-electron interaction. The complex QD dynamics has been reduced to the
magnetization field dynamics governed by the effective Keldysh action being a
nonlinear functional of the magnetization field. We have expanded this action
up to the second order in the magnetization field. This expansion represents
an expansion about the weak coupling fixed point and thus must reproduce the
unitary limit of weakly correlated QDs. The QD TDOS has been derived and the
differential conductance has been calculated as a function of the temperature
and voltage. These calculations have correctly reproduced the conductance
maximum and thus confirmed the consistency of our theory establishing an
alternative versatile and simple tool to explore nonequilibrium weakly
interacting QDs, in particular, in the unitary limit which becomes more and
more relevant in modern experiments.
\\
\section{Acknowledgment}
Support from the DFG SFB 689 is acknowledged.

\end{document}